\pdfoutput=1

\documentclass[10pt,conference]{IEEEtran}
\usepackage[outdir=./]{epstopdf}

\usepackage{amsmath}
\usepackage{subfig}
\usepackage{graphicx}
\usepackage{hhline}
\usepackage[compress]{cite}
\usepackage{textcomp}
\usepackage{pifont}
\usepackage{tabularx}
\usepackage{cite}
\usepackage{tikz}
\usepackage{fancyhdr}
\usepackage[absolute,showboxes]{textpos}

\ifCLASSINFOpdf
\else
\fi
\TPMargin{5pt}

\IEEEoverridecommandlockouts
\begin{document}

\title{Towards Policy Enforcement Point as a Service (PEPS)\thanks{This is a copy of the paper accepted at IEEE NFV-SDN'16. An extended work based on this paper will be submitted to a journal.}}

\author{
\IEEEauthorblockN{Arash Shaghaghi\textsuperscript{1,2}, Mohamed Ali (Dali) Kaafar\textsuperscript{2}, Sandra Scott-Hayward\textsuperscript{3}, Salil S. Kanhere\textsuperscript{1} and Sanjay Jha\textsuperscript{1}}
\IEEEauthorblockA{\textsuperscript{1}School of Computer Science and Engineering, UNSW Australia, Sydney, Australia \\
\textsuperscript{2}Data61, CSIRO, Australia \\
\textsuperscript{3}Centre for Secure Information Technologies (CSIT), Queen's University Belfast, Northern Ireland \\
Contact: a.shaghaghi@unsw.edu.au}}


%Telephone: (800) 555--1212, Fax: (888) 555--1212}

%\IEEEoverridecommandlockouts
%\IEEEpubid{\makebox[\columnwidth]{978-1-5090-0933-6/16/\$31.00~\copyright~2016 IEEE \hfill} \hspace{\columnsep}\makebox[\columnwidth]{}}

\maketitle



% \fi
\IEEEpeerreviewmaketitle

\begin{abstract}
In this paper, we coin the term Policy Enforcement as a Service (PEPS), which enables the provision of innovative inter-layer and inter-domain Access Control. We leverage the architecture of Software-Defined-Network (SDN) to introduce a common network-level enforcement point, which is made available to a range of access control systems. With our PEPS model, it is possible to have a `defense in depth' protection model and drop unsuccessful access requests before engaging the data provider (e.g. a database system). Moreover, the current implementation of access control within the `trusted' perimeter of an organization is no longer a restriction so that the potential for novel, distributed and cooperative security services can be realized. We conduct an analysis of the security requirements and technical challenges for implementing Policy Enforcement as a Service. To illustrate the benefits of our proposal in practice, we include a report on our prototype PEPS-enabled location-based access control.
\end{abstract}

\section{Introduction}
With Software-Defined-Network (SDN), the separation of control and data plane and programmability in the network enable provision of enhanced security systems. A diverse set of proposals have emerged that exploit the architecture of SDN, and specifically the network-wide view of SDN controllers, to implement reactive monitoring and automated response systems. Recently, an emerging body of literature is shaped around the idea of using SDN to introduce innovative security services. We follow the latter approach and leverage the capabilities of SDN in moving towards a new model of access control enforcement, which could potentially open the door to a range of new types of security services. \par

\begin{figure}[!t]
\centering
\includegraphics[width=3.65in, height=2.2in]{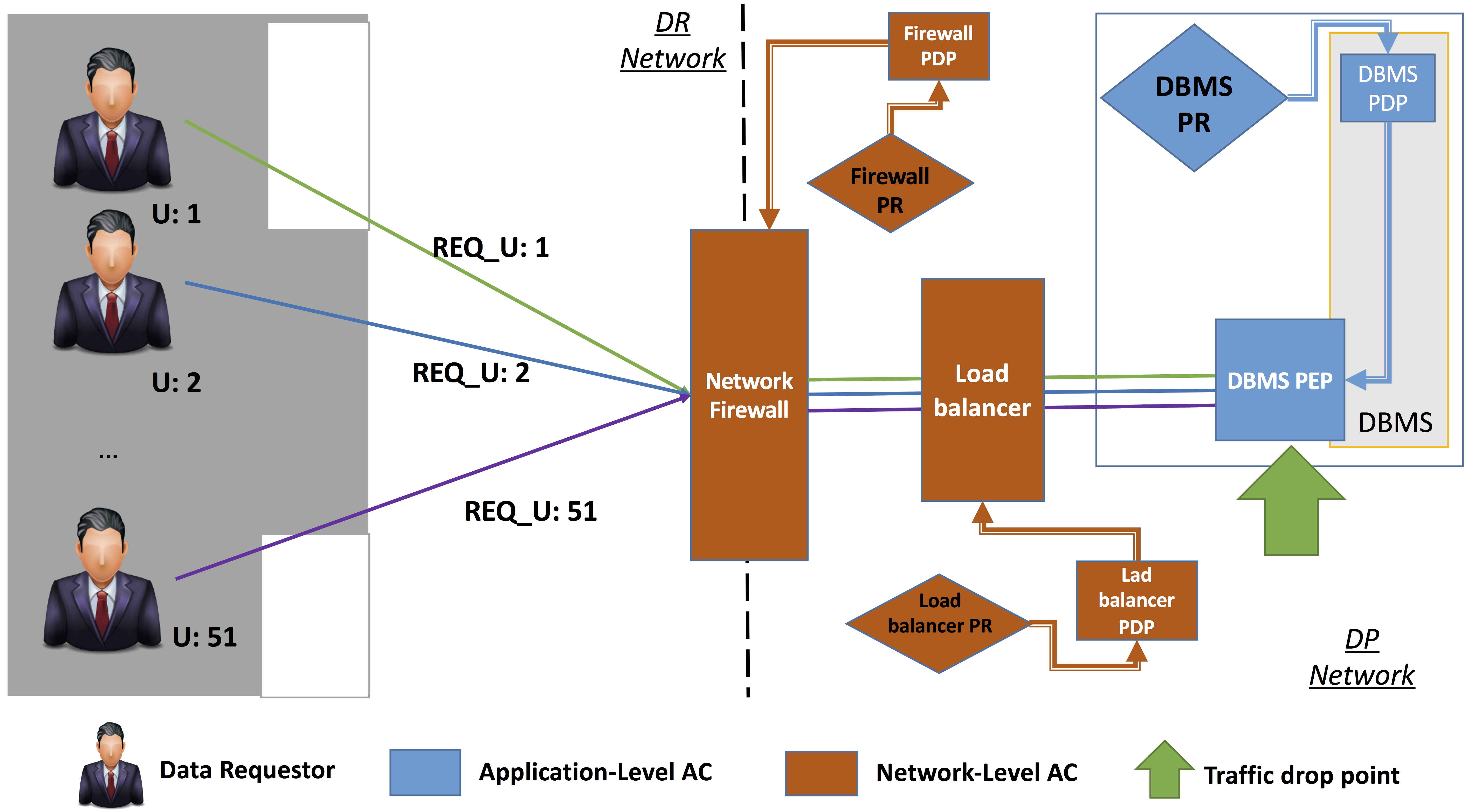}
% where an .eps filename suffix will be assumed under latex, 
% and a .pdf suffix will be assumed for pdflatex; or what has been declared
% via \DeclareGraphicsExtensions.
\caption{A typical access control process flow between a Data Provider (DP) and Data Requestor (DR) located in separate networks.}
\label{fig_typical}
\vspace{-1.5em}
\end{figure}

Access control systems limit the operations of legitimate users \cite{sandhu1}. The main components of an access control system include Policy Decision Point (PDP), Policy Repository (PR) and Policy Enforcement Point (PEP). Accordingly, an authorization flow involves retrieving the user access request by PDP, inquiry the PR for matching policies and enforcing the decision by PEP. Figure \ref{fig_typical} illustrates a \textit{typical} access control process flow between a Database Management System (DBMS), as the Data Provider (DP), and a user at a remote network, as the Data Requestor (DR). An access request by a DR is sent from the DR network to the DP network, where the DBMS makes the access decisions and enforces them. In other words, with this setup, an access request reaches DR at application-layer and only then is decided about. Hence, an attacker is allowed to engage the system and its hosting network and possibly execute certain types of attacks such as Denial of Service (DoS) or port scanning. 

%Currently, the PEP of access control systems is placed within the `trusted' perimeter of an organization and each system relies on its own independent mechanism. For example, when enforcing restrictions to flows passing through the switches under its governance, an SDN firewall application operates independently of access control components of web applications, file servers, delegation systems and etc. Similarly, access requests sent to a database system are retrieved by the Database Management System (DBMS) and authroized according to its own independent Policy Repository (PR). \par

In this paper, we propose to leverage the capabilities brought by SDN to introduce programmable network-level policy enforcement points, which application-layer services may subscribe to. The extra enforcement points serve to create a `defense in depth' \cite[p.~308]{anderson2008security} model of protection and improve the protection of services hosted in enterprise-like networks. PEPS enables applications such as DBMS to enforce dynamic access control policies both at a lower-level (i.e. network-level enforcement rather than application-level) and closer to the DR's network (i.e. inter-domain enforcement). In effect, PEPS enables authorized system resources to push pre-approved policies to a purpose-built SDN application, which enforces these policies at the level of SDN switches. We coin Policy Enforcement Point as a Service (PEPS) for this model of enforcement.

Referring to Figure \ref{fig_typical}, with PEPS, instead of waiting for the requests to reach the DBMS's PEP, the DBMS may instruct the network to drop requests originated from a specific network address for a certain period. Similarly, for Quality of Service (QoS) purposes the DBMS may instruct the firewall to adjust traffic volume forwarded to it. Moreover, if the two network-domains were to collaborate, the DBMS may push dynamic and pre-approved policies to the DR's network and block unauthorized access requests either pro-actively or reactively. For example, access requests from `non-secure' areas of a building destined to the DBMS may be dropped as early as entering the DR's network. We remind that in defense in depth model of protection, the outer-layer defenses may be less reliable than the inner-layers. Hence, if, for any reason, the DR's network fails to ensure to the remote policies, the standard DR's PEP is still in effect.

%PEPS may have ad-hoc and manual equivalent in traditional networks using network middle-boxes but it may lead to network performance bottle-necks and cause conflicts with multiple-services (local or remote) subscribing to it. With SDN, however, 

The resulting protection with PEPS is significantly different and novel compared to status-quo. In fact, from an access control viewpoint, the extra enforcement points at SDN's data plane, facilitates moving towards distributed and cooperative enforcement of access control for application and services. PEPS also motivates a new line of thought in access control, which is deploying verifiable protection points beyond the trusted perimeter of an organization. \par

The rest of this paper is structured as follows. In Section \S\ref{sec:background} we briefly revise background information on Access Control and SDN security. Thereafter, in \S\ref{sec:peps}, we elaborate on our motivation and preliminary technical requirements for implementing PEPS. In \S\ref{sec:usecase}, we report on our prototype implementation of a PEPS-enabled location-based access control (LBAC) system. The advantages of our LBAC compared to state-of-the-art is discussed to motivate further investigation of various applications of PEPS. We conclude this paper specifying our work-in-progress and outlining suggestions for future work.

%Such dynamic application-defined network-level enforcement results in having a new set of network policies that has not been considered before.

\section{Background} \label{sec:background}
\subsection{Access Control} \label{subsec:ac}
Every user's attempt to interact with protected resources is mediated by access control - the oldest information security mechanisms. During the last decade, an increasing number of major data leakage incidents are associated with the failure of access control \cite{harvard}. Security researchers \cite{lee2011advances, rajarajan2012security, park2004ucon}, associate this to the incompatibility of currently implementable access control with today's requirements. Hence, an increasing number of researchers are investigating innovative proposals to change this condition \cite{YvoArash}. One of the promising directions is the interaction of access control with other security services. For example, Crampton et al. propose integrating intrusion detection systems with access control systems \cite{crampton2010towards}. \par

Distributed access control is a fairly recent trend in access control. For example, in \cite{tsankov2014decentralized}, authors propose having multiple principals defining the policies for PDP. Nevertheless, the enforcement is through a single trusted reference monitor. Digital Rights Management (DRM) \cite{subramanya2006digital} is another example, which is constituted of distributed enforcement. With DRM, the client-side enforcement is, in fact, an extra point of enforcement that facilitates a more granular control over information. DRM is well-recongized and appreciated by industry, and its architecture has been inspiring for our work. \par

\subsection{Software-Defined-Network Security} \sloppy \label{subsec:sdn}
SDN Security literature may be split into two main categories, securing the Software-Defined-Network itself or leveraging the capabilities of this technology for security services. In \cite{scott2015survey}, Scott-Hayward et al. provide a categorization of the security issues associated with the SDN framework, and detail the body of literature focussed on solutions to these threats. The security requirements of PEPS defined in \S3.3 rely on such solutions. \par
On the other hand, SDN facilitates the provision of reactive and automated monitoring, analysis and response systems. The key SDN characteristics contributing here are the network-wide view for centralized monitoring \cite{alsmadi2015security} and the programmability of SDN to redirect selected network traffic through middleboxes (see \cite{anwer2013slick}, \cite{fayazbakhsh2013flowtags}, and \cite{qazi2013simple} for examples). Along with the improvement of traditional security solutions via SDN, novel security services are also built on top of SDN. For example, \cite{hassan} uses SDN to develop an architecture that enables residential internet customization, which could be used to secure household appliances. \cite{mehdi2011revisiting} and \cite{shin2012cloudwatcher} also introduce innovative services. \par
Recently, a few number of solutions extend the Authentication, Authorization, and Accounting (AAA) functionality using the SDN controller and focus on identity management and authentication mechanisms (e.g. \cite{mattosauthflow} and \cite{dangovas2014sdn}, \cite{toseef2014c}). Our PEPS model is a network-level access control implementation deployed at the SDN data plane.
%In this paper we focus on the PEP component of access control, which has rarely been revised since its initial conception. \par
 \section{Policy Enforcement as a Service} \label{sec:peps}
\subsection{Motivation} \label{subsec:motivation}
Every organization has a number of systems equipped with their own access control mechanism, e.g. file systems, firewalls, location-detection, etc. The access control component of these systems operates independently. Hence, if any of these PEP fail then unauthorized access to data is inevitable. As mentioned in \ref{subsec:ac}, distributed reference monitors have been previously investigated in the literature. However, to the best of our knowledge, the idea of having a cooperation among PEP has not yet been explored. Recalling that in most cases access requests to data, or resources, are mediated through the network we believe it is possible to place a shared enforcement point for all services to use. However, unlike firewalls, this component has to adhere to dynamic policies and requirements of application-layer systems.  \par

Moreover, letting applications such as DBMS instruct the network may result in better and more dynamic network management. For example, assume at time t of day d the network infrastructure hosting the DBMS is congested and can only handle 50 concurrent connections to DBMS due to the global QoS requirements. Accordingly, the DBMS administrator defines a policy to drop connection requests beyond 50 and instructs the DBMS'PEP to limit the total number of requests from a single source to 10. The issue with this arrangement is that the UNSW network Admin has to trust the DB Admin and the DBMS access control for this as such temporary policies are application-dependent and are unknown to the network components such as a firewall. Furthermore, with application-level access control traffic still reaches the network and attacks such as DoS may still target the network hosting the DBMS.

Thirdly, dropping traffic associated with unauthorized requests closer to the source would enable saving significant traffic from flowing over the networks or Internet.
\subsection{Proposed Approach}
We propose designing a shareable enforcement point at network-level, which is made available to application-layer access control systems. The shareable enforcement point is made available as a service and application or services need to subscribe to use it. We coin the term `Policy Enforcement as a Service', or PEPS, for this security service. \par

Relying on traditional networks and deploying middle-boxes for PEPS would be challenging. Specifically, policy conflict resolution and performance management will be inefficient and troublesome. However, the SDN architecture is well-suited for such requirements since the controller composes policies received by various applications and there is an on-going effort to optimize this process with respect to dynamic and reactive policies. \par

In SDN, the control plane entails both PR and PDP and the data plane is equivalent to PEP in access control. In essence, the SDN controller takes as input an extra set of policy for PEPS, which may be defined by local or remote application-layer access control systems. We design an SDN application responsible to retrieve these policies and submitting them to the network operating system.

\subsection{Assumptions}
We require the following assumptions to hold:
\begin{itemize}
\setlength\itemsep{0.1em}
\item The SDN controller and external SDN applications are assumed to be secure and able to communicate securely (e.g. using TLS).
\item The SDN data plane is not compromised.
\item The east and west bound communication link between controllers in different networking domains is secure.
\end{itemize}
As mentioned in \ref{subsec:sdn}, there is an over-expanding body of literature exploring the security of SDN both at data plane and control plane. Similar to various proposals that leverage SDN to introduce novel services and applications (see \S\ref{subsec:sdn}), we focus on our proposed system assuming the underlying platform is reasonably reliable and secure.

\subsection{Security Requirements} \label{subsub:threat}                           
A PEPS solution should be designed and implemented such that a malicious subscriber, whether in the same perimeter or not, \underline{cannot}:
\begin{itemize}
\setlength\itemsep{0.1em}
\item Violate the policy specifications of the service provider through the remote policies.
\item Violate the policy specifications of other services, which use the enforcement point, whether in the same perimeter or not.
\item Affect the performance of the SDN controller itself. For example, causing a DoS attack with constant update of the remote policies.
\end{itemize}
%\item PEPS must not become a bottleneck for the networking performance.  

\subsection{Main Components and Requirements} 

\begin{figure}[!t]
\centering
\includegraphics[width=3.5in, height=1.6in]{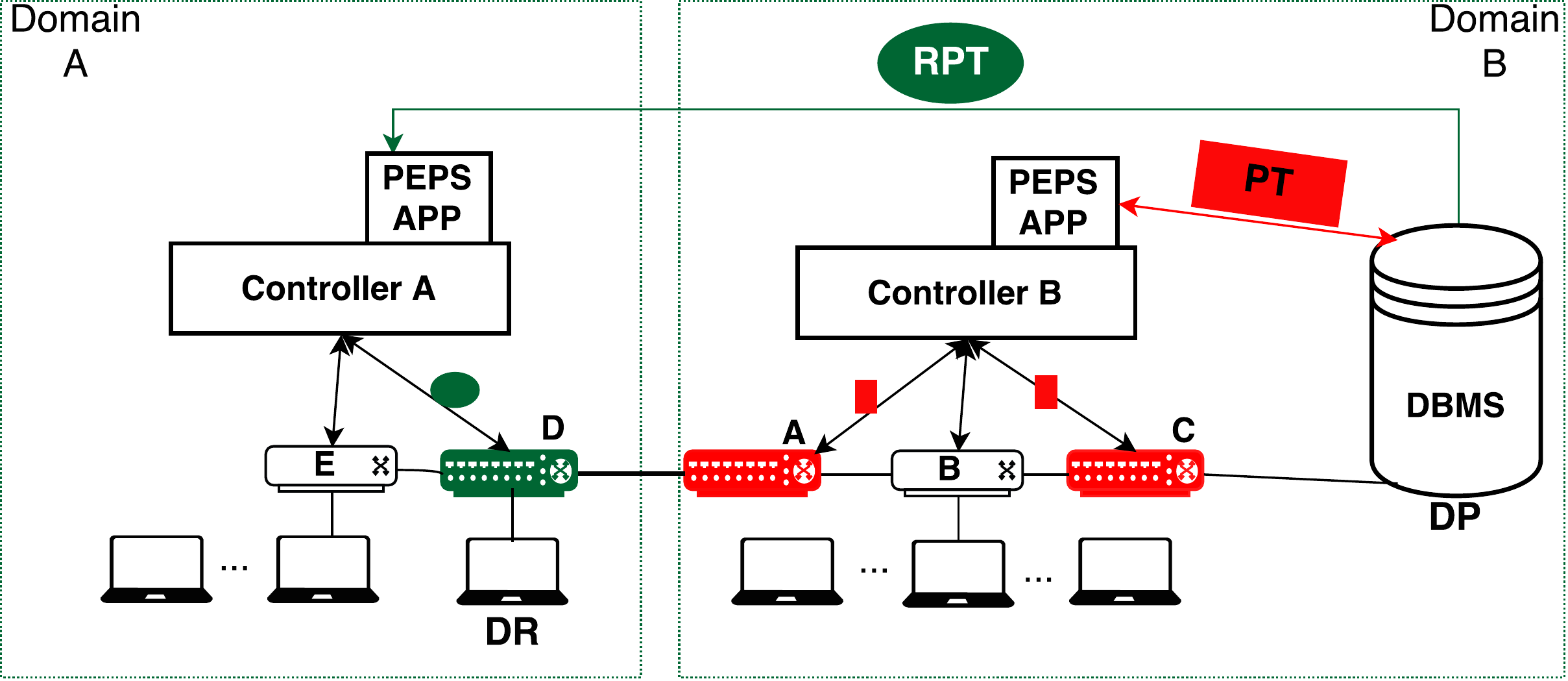}
% where an .eps filename suffix will be assumed under latex, 
% and a .pdf suffix will be assumed for pdflatex; or what has been declared
% via \DeclareGraphicsExtensions.
\caption{Abstract representation of Policy Transfer (PT) and Remote Policy Transfer (RPT) in SDN networks deploying PEPS. Switches in red and gree colour are effected by PT and RPT, respectively.}
\vspace{-1.5em}
\label{fig_architecture}
\end{figure}

Figure \ref{fig_architecture} shows the main components required in an SDN network deploying PEPS.
\textbf{Policy Transfer} is the standard protocol used to define policies at application-layer (e.g. by DBMS) for network-level SDN application. Similarly, \textbf{Remote Policy Transfer} is used to translate application-layer policies for a remotely located SDN network deploying PEPS. RPT is securely exchanged over east and westbound link between controllers and PT is exchanged over a secure connection.

Conflicting policies will result in one or more of the threats mentioned in \ref{subsub:threat}. Therefore, we have to ensure the following three requirements are met:

\textbf{Requirement 1}: \label{condition1}
    Let $P'$ be the set of policies for controller $C_{1}$, which is in domain $D_{1}$ and governs over the set of switches $S$. We define $P_{r}$ as the PT for $C_{1}$ and say: $P_{i}$ is a valid PT for $P'$ if and only if $P = \{P' \cup P_{r} \}$ does not violate the original policy specification $P'$.

\textbf{Requirement 2}: ensures the remote policies do not conflict with original policy specification. Therefore, we just replace PT with RPT in Requirement 1.
  
%In order to ensure that Requirement \ref{condition1} is met, we use Remote Policy Template (RPT). RPT is an XML-like file, which is human readable, and is used as a standard for defining the access control requirements. RPT is then converted into flow-table rules for a segment of the network. The templates are defined by the network administrator in different categories, e.g. for beyond-perimeter services and within-perimeter services. An RPT, in the simplest case includes Match, Action and Description tags. \par
  
Policy composition and conflict detection is an ongoing challenge in Software-Defined-Network \cite{kreutz2015software}. In order to prevent adding further complications to this domain with PEPS, it is best to restrict the capabilities of RPT at this time. We postulate to restrict a PEPS service subscriber only to submit RPT that relate to flow destined directly towards it (e.g. DB in Domain B may only set RPT at domain A for traffic flowing towards it's own domain). Moreover, the priority of rules set after conversion of RPT should always be set below any matching policy set locally. Accordingly, we define Requirement 3:

\textbf{Requirement 3}: \label{condition2}
    Let $P'$ be the set of policies for controller $C_{1}$, which is in domain $D_{1}$ and governs over the set of switches $S$ and \textit{has been defined locally}. We define $P_{r}$ as the remote policy for $C_{1}$, which is generated according to RPT. Then, having $\exists P_{r_{i}}$  that  $\bot$ $P'_{i}$ results in $P'_{i}$ OVERRIDES $P_{r_{i}}$ in the final policy set $P = \{P' \cup P_{r} \}$. 

\subsection{Practical Considerations}

\textbf{Multi-Table Pipeline:} the data plane of SDN supports Flow Table Pipeline (FTP) - introduced with OpenFlow specification V1.1 to improve the flow processing performance \cite{openflow1}. The pipeline consists of multiple flow tables. The incoming packet is first matched with the first flow table, where the specified actions could direct the packet to another flow table for further processing of the packet. With this redirection mechanism, the SDN control plane could build a logical single source directed acyclic graph on the FTP for processing. \par
To implement non-conflicting remote policies we propose customized use of FTP. All flow rules resulting from PT or RPT should be added to the last flow table. This flow table is directly managed by our purpose built PEPS APP. The incoming flow to the switch is first-matched against all but the last flow table (i.e. rules required by local policies are first processed), and if a flow is still allowed, then it is passed to the final flow table for processing. In other words,

Let $FTP$ be a set of flow tables $\{FT_{1}, FT_{2}, ..., FT_{n}\}$, $FT_{i}$ for $i < n$ generated according to the set of policies $P'$ for Controller $C_{1}$, $FT_{n}$ set according to remote policy $P_{r}$ for $C_{1}$. Then, an incoming packet $Pckt$ is MATCHED against $FT_{i}$ for $i<n-1$. The resulting $Pckt'$ is then MATCHED against $FT_{n}$.

This simplifies conflict resolution between local and remote policies when using FTP.

\textbf{Multiple PEPS SDN Application Instances:} PEPS APP is installed on networks deploying PEPS model of enforcement. This application is responsible to retrieve PT and RPT and to convert them into flow-table rules for submission to the controller. PEPS should be securely connected to application-layer services sending PT or RPT. Moreover, we must ensure PEPS has minimum impact on the controller performance. Network-Function-Virtualization (NFV) may be used to improve the PEPS performance.

\section{PEPS in Practice} \label{sec:usecase}
We now report on our prototype implementation of a PEPS-enabled location-based access control. This section aims to highlight the advantages of PEPS in practice and motivate future work.

Location-based access controls rely on user's location as one of the attributes when making access decisions. There are simple solutions to retrieve user's location. For example, it is possible to retrieve user's location using the device integrated peripherals such as GPS device. However, proof of presence is a challenging aspect of location-based services, especially for an indoor environment. As thoroughly discussed in \cite{miettinen2015know}, proof of presence schemes can be categorized into beaconing-based, context-based and distance-bounding based approaches. 
%The first category are schemes that are based on beaconing of information into the context such as Geobadges and NFC-based solutions. The second category includes schemes such as co-location verification and using device sensors such as ambient light to fingerprint the context of a location. The last category requires special hardware to perform distance bounding. \par
Most of the proof of presence solutions are challenged for one or more of the following reasons: requiring specialized hardware or software, being immobile, unable to track movement in real-time (or requiring extensive ongoing context scans either by Data Provider or Data Requestor), being computationally hard or infeasible, or being extremely privacy-invasive. Hence, in practice, the adoption of these schemes by organizations is challenging (e.g. \cite{luo2010proving}, \cite{wu2006challenges}). \par
Here, we propose and implement two alternative approaches to ensure proof of presence and enforce location-based access control using PEPS model. These schemes are not originally built to replace existing solutions. Instead, we are interested to use them as the first layer of defense (i.e. the outer layer of defense in depth model). We define a scenario in which there are two organizations both with SDN networks. The Data Provider (DP) resides in network B, and the Data Requestor (DR) is located in network A. We have implemented the following scheme within a simulated environment using Mininet 2.2.0 and Floodlight V.1 running as the SDN controllers. The applications have been developed for this controller and communicate over a secure TLS connection with an open source database server, MariaDB, as the Data Provider. We have integrated an extra module into MariaDB, which mediates communication and coordinates with SDN PEPS APP both in the local and remote networks. \\

\textbf{PEPS-enabled location-based access control with real-time location tracking} \label{subsub:rt}

\begin{figure}[!t]
\centering
\includegraphics[width=3in]{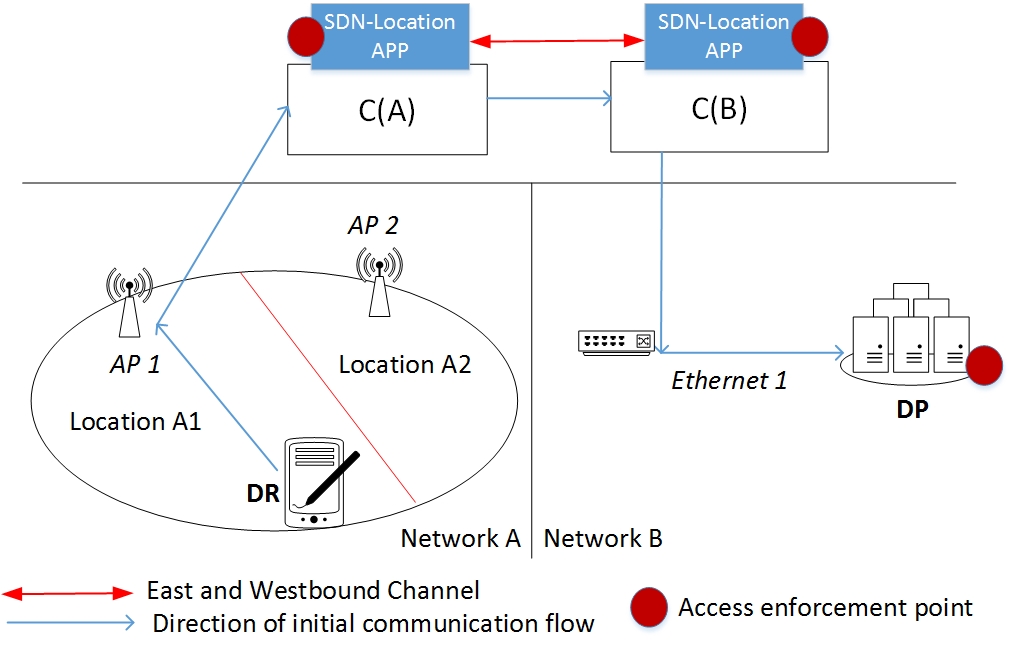}
% where an .eps filename suffix will be assumed under latex, 
% and a .pdf suffix will be assumed for pdflatex; or what has been declared
% via \DeclareGraphicsExtensions.
\caption{Policy enforcement points that exist with PEPS are depicted within a simplified location-based access control. Without PEPS, the only PEP would be at DP.}
\vspace{-0.5em}
\label{fig_3}
\end{figure}

\textit{SDN-based location tracking:} we use OpenFlow to retrieve the location of users in real-time. This is a new approach to track users and can be easily deployed without any specialized hardware in SDN networks. Whenever a packet is received by a switch, and it does not match any of its existing forwarding rules then a \textit{packet\_in} message containing the \textit{switch ID} and \textit{port ID} is sent to the governing controller. The controller uses this information to create a dynamic geo-location lookup table. This table matches the user's device IP to a switch port. The network locations retrieved through \textit{switch ID} can be matched to different sections within the building. For example, in Figure \ref{fig_3}, \textit{Location 1} is associated to \textit{AP 1}. An issue to consider for wireless devices would be managing the signal coverage that could mislead this scheme. This can be solved using proper and careful positioning of these devices and signal blocking solutions \cite{coleman2010certified}. Indeed, the cost of performing such is much lower than having specialized equipment for location detection. Moreover, an important advantage of this scheme is that unlike most proof of presence schemes, it is capable of tracking the movement of the user around the locations in real-time. It is possible to ensure that this scheme is secure against IP Spoofing by setting a rule that only packets from a specific IP address are forwarded from the switch port. \par
\textit{PEPS-based Access Enforcement:} at this point, using the above scheme, we build a location-based access control model on top of our PEPS model. As depicted Figure \ref{fig_3}, we require an \textit{SDN-Location App} (equivalent to PEPS APP referred to earlier) installed on both DP and DR networks. An RPT, issued by the DP, defines that any traffic destined to DP is dropped unless the SDN-Location APP on the requesting side initiates a valid session with the same application on the provider side. A valid session requires that the user requesting data be located by the \textit{SDN-Location App} and is allowed to communicate with DP in accordance with the rules extracted from RPT. Only then a host is allowed to send a request for data. As also depicted in Figure 3, compared to existing approached, with our location-based access control model there are extra network-level enforcement points both at source network and host. \\
%Here, the extra service defined for MariaDB is only used to send specific \textit{block\_access} request to its own \textit{SDN-Location App}, when and where needed. The SDN-Location App could then send an updated RPT and also block traffic itself. So, considering that the database server also has its own enforcement point, we have three layers of blocking the unwanted traffic (See Figure \ref{fig_3}). \par

\begin{figure}[!t]
\centering
\includegraphics[width=3.5in]{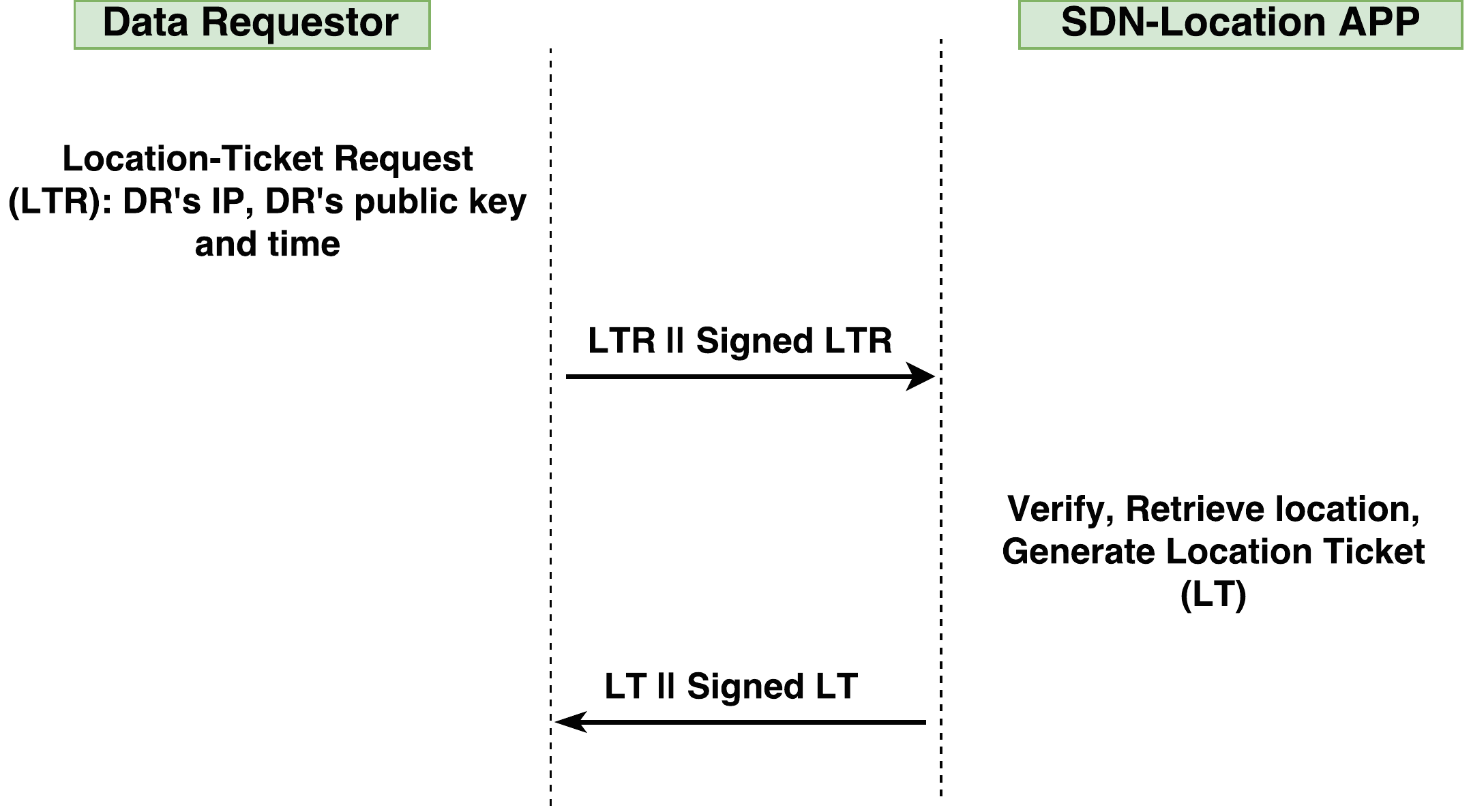}
% where an .eps filename suffix will be assumed under latex, 
% and a .pdf suffix will be assumed for pdflatex; or what has been declared
% via \DeclareGraphicsExtensions.
\caption{Representation of proposed ticketing protocol.}
\label{fig_2}
\vspace{-1em}
\end{figure}

\textbf{PEPS-enabled location-based access control with location-tickets} \label{sub:locationtrack} 

\begin{figure}[!t]
\centering
\includegraphics[width=3.5in, height=1.5in]{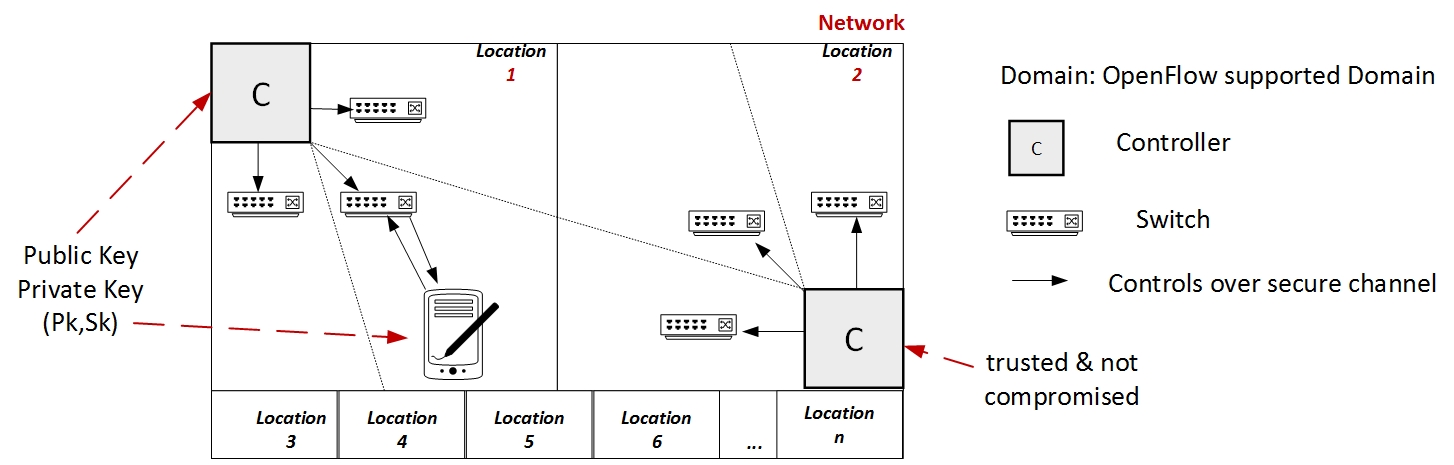}
\caption{Assumptions and requirements for the location-ticket (LT) scheme.}
\label{fig_1}
\vspace{-1.5em}
\end{figure}

\textit{The SDN-based Location Ticketing Scheme:} it is possible to use the same location detection scheme to generate location tickets - rather than real-time tracking. The assumptions and requirements for the location-ticket scheme is depicted in Figure \ref{fig_1}. Each controller and user are equipped with a public and private key. The DR creates a Location Ticket Request \textit{LTR} containing the DR's IP address, public key and time. It digitally signs \textit{LTR} and sends it to the \textit{SDN-Location App} running on top of the controller. The signature is verified, and the IP address is compared with the one in the packet header. If the IP is legitimate, the user's location is retrieved using the same approach mechanism described earlier. A Location Ticket (LT) is then generated using the DR's IP address, its public key, time and location. LT is signed and sent along with LT to the DR. The protocol is represented in Figure 4. 

The proposed location ticket scheme binds the DR's IP and public key together. This helps to prevent one of the main threats against proof of presence schemes such as Sybil Attack, where users create several fake identities in several locations within the network.

\textit{PEPS-based Access Enforcement:} the location-ticket scheme facilitates the integration of PEPS with existing application and services. Specifically, unlike the real-time approach, there is no requirement of having SDN APP on both DP and DR. A location ticket issued by SDN APP at DR may be provided to any application or service requesting proof of presence. The LT scheme also removes the requirement of session establishment between remote controllers, which may be more practical in many scenarios. We implemented the LT scheme and sent location tickets along with access requests to MariaDB as part of our prototype implementation.

\subsection{Security and Performance Analysis}
\textbf{Performance Analysis:} we simulated a network with 32 switches and four threads and sent location ticket requests to the application running on top of the controller. Figure 5.a shows the standard performance of the Floodlight controller when not running the \textit{SDN-Location App}. We then ran the application and issued 1000 LTR. The controller performance was steady and cumulative distribution function (CFD) showed reasonable performance impact. However, as we increased the LTR numbers the performance of the controller when handling incoming flows degraded --- compare Figure 5.b with 5.a. This points us to the fact that it may be a better approach to outsource demanding processes and use solutions such as NFV. \par
\textbf{Security Analysis:} we include an analysis of SDN-based location detection scheme. The security and performance of PEPS is included in Section 5. \par
The scheme does not rely on user's device peripherals and is built on capabilities available at network infrastructure level. Hence, it is much harder for an attacker to compromise the system. Also, since this scheme does not rely on context measurement information, it is secure against most recent attacks including Context Guessing Attack \cite{miettinen2015know}. Moreover, this scheme could be used as a standalone solution --- not for proof of presence but actual location detection. If so, it allows the protection of user's privacy against service providers that retrieve a huge amount of personal information when retrieving the device location. However, the original scheme is vulnerable to the Wormhole attack. It is possible to solve this problem using authenticated Ping and various other network delay measurement techniques. As further security analysis and improvement is beyond the scope of this paper and we leave this for our future work.
\begin{figure}[!t]
\centering
\includegraphics[width=3in, height=1.5in]{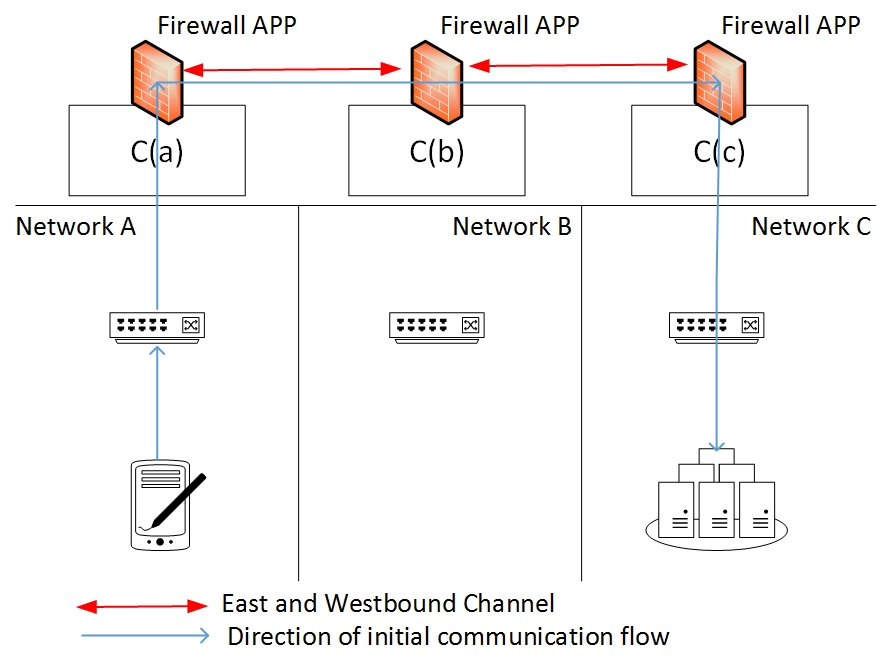}
% where an .eps filename suffix will be assumed under latex, 
% and a .pdf suffix will be assumed for pdflatex; or what has been declared
% via \DeclareGraphicsExtensions.
\caption{Abstract representation of progressive layered firewall model.}
\vspace{-1.5em}
\label{fig_4}
\end{figure}

\section{Discussion} \label{sec:discussion}
As illustrated in Figure \ref{fig_3}, our PEPS-based access control model allows having a defense in depth model of protection. This change in access control enforcement has several advantages. For example, it allows network bandwidth to be saved by blocking unauthroized requests at the source. It also enables s of certain categories of attacks, where the attack is based on challenge and response (e.g Port Scanning). Evidently, dropping traffic before engaging services or systems also facilitates protection against DoS threats. 

PEPS enables having a more context-aware access control. For example, if the remote enforcement is not blocking traffic as expected then it could be considered as less trustworthy. Accordingly, if controllers in different domains were to share knowledge about this, they could block all, or specific, access requests originated from the suspicious network until further investigation (e.g. the controller may be compromised or the PEPS APP may be malfunctioning). \par

We presume the aforementioned are only some of the advantages of brought with a PEPS model of access control enforcement. Specifically, the co-operation of domains in access control could lead developing novel security services never sought before. For example, we are investigating the development of a PEPS-enabled inter-domain firewall system, which gradually and progressively applies policies (see Figure \ref{fig_4} for an abstract representation). In other words, the RPT mechanism used to define non-conflicting remote policies could be used between firewall applications of SDN controllers to progressively block unwanted traffic reaching an organization network. It should be noted that, from a practical point of view, such approach may not have been feasible with existing firewall solutions without SDN and conceptualization of PEPS. For example, firewalls may have been from different providers and cooperation would not have been feasible. We leave further investigation and exploration as future work. 
\begin{figure}[ht!]
\centering
\subfloat[][]{\includegraphics[width=.26\textwidth]{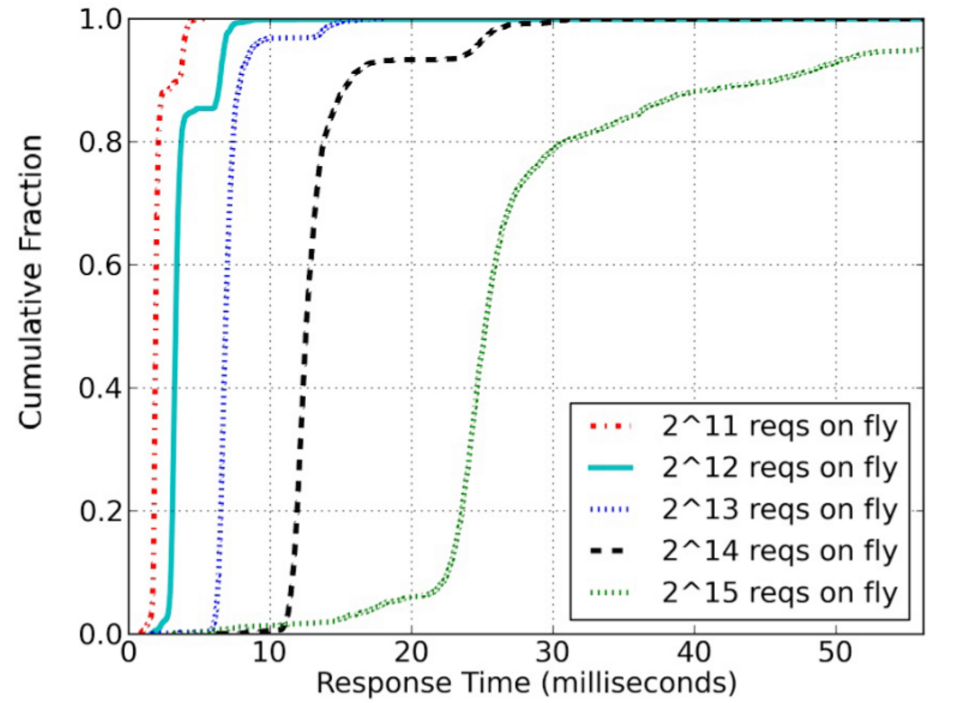}\label{iter1000}}
\subfloat[][]{\includegraphics[width=.24\textwidth]{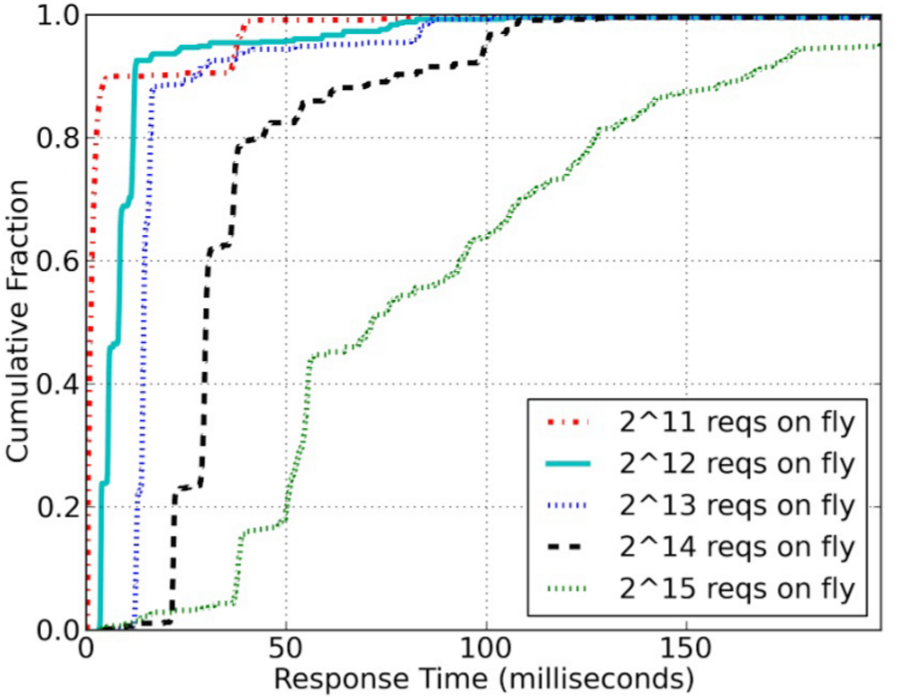}\label{iter4000}}\\
\caption{Impact of SDN-Location APP on Floodlight controller. Figure (a) is without the application running and Figure (b) is with the application running.}
\vspace{-1em}
\label{fig:leakagethroughputALL}
\end{figure}

PEPS is currently at its conception phase and requires much further exploration and development before coming into practice. Specifically, the translation of PT and RPT for the network hosting PEPS is a challenging issue -- e.g. which forwarding devices will have to apply the remote policies in the network. Moreover, the impacts of PEPS on network performance and security threats associated with it require proper analysis. We remind that our early performance evaluation is not prohibitive (see \S\ref{sec:usecase}). 

\section{Conclusion and Future Work}
In this paper, we revisited the Policy Enforcement Point (PEP) of access control. We introduced Policy Enforcement Point as a Service, or PEPS, by leveraging the capabilities of Software-Defined-Network (SDN). PEPS allows cooperation of PEP among application-layer and network-layer services either in the same network or remote domains. It enables improving the security of application-layer services hosted in networks and promises the development of innovative collaborative network-based security services. Beyond conceptualization, we made an early attempt to discuss practical requirements for PEPS and reported on our prototype implementation. Detailed analysis of some of the security challenges of PEPS and a more technical exploration on how to integrate remote policy is left as our future work. 

%\end{document}  % This is where a 'short' article might terminate

%ACKNOWLEDGMENTS are optional
%\section{Acknowledgments}
%SOME TEXT.

\bibliographystyle{abbrv}
\bibliography{sigproc}  % sigproc.bib is the name of the Bibliography in this case

% ACM needs 'a single self-contained file'!
%\balancecolumns
\end{document}